\documentclass[sigconf]{acmart}
\usepackage{textcomp}
\usepackage{graphicx} 
\usepackage{comment}
\usepackage{xcolor}
\usepackage{enumitem}
\usepackage{amsmath}
\usepackage{setspace}
\usepackage{array,multirow}
\usepackage{longtable}
\usepackage{etoolbox} \usepackage[normalem]{ulem}
\usepackage{flushend}
\usepackage[absolute,overlay]{textpos}
\AtBeginDocument{  \providecommand\BibTeX{{    \normalfont B\kern-0.5em{\scshape i\kern-0.25em b}\kern-0.8em\TeX}}}

\renewcommand\footnotetextcopyrightpermission[1]{} % removes footnote with conference information in first column
\newtoggle{comments}

\newcommand{\fakeheader}[1]
{\vspace{.75mm} \noindent\textbf{#1}~}

\toggletrue{comments}

\iftoggle{comments}{
    \newcommand{\notecn}[1]
    	{{\color{teal}[{\bf Chris:} #1]}}
    \newcommand{\notejs}[1]
    	{{\color{red}[{\bf Jat:} #1]}}
    \newcommand{\noterc}[1]
    	{{\color{magenta}[{\bf Richard:} #1]}}
    \newcommand{\noteaj}[1]
    	{{\color{blue}[{\bf Ahmad:} #1]}}
    \newcommand{\notejc}[1]
    	{{\color{purple}[{\bf Jennifer:} #1]}}
    \newcommand{\todo}[1]
    	{{\color{red}[{\bf Todo:} #1]}}
    \newcommand{\cut}[1]
    	{{\color{red}\sout{#1}}}

}{
    \newcommand{\notecn}[1]{}
    \newcommand{\notejs}[1]{}
    \newcommand{\noterc}[1]{}
    \newcommand{\noteaj}[1]{}
    \newcommand{\notejc}[1]{}
    \newcommand{\todo}[1]{}
    \newcommand{\cut}[1]{}

}

\begin{document}
\fancyhead{}

\title{Monitoring Misuse for Accountable\\`Artificial Intelligence as a Service'}

\author{Seyyed Ahmad Javadi, Richard Cloete, Jennifer Cobbe, Michelle Seng Ah Lee and Jatinder Singh}
\affiliation{Compliant \& Accountable Systems Group, Dept. of Computer Science \& Technology\\University of Cambridge, UK}
\email{firstname(s).lastname@cst.cam.ac.uk}

\renewcommand{\shortauthors}{S.A. Javadi, et al.}

\begin{abstract}
 AI is increasingly being offered `as a service' (AIaaS). This entails service
providers offering customers access to pre-built AI models and services, for tasks such as object recognition, text translation, text-to-voice conversion, and facial recognition, to name a few.  
The offerings enable customers to easily integrate a range of powerful AI-driven capabilities into their applications. 
Customers access these models through the provider's APIs, sending particular data to which models are applied, the results of which returned.

However, there are many situations in which the use of AI can be problematic.
AIaaS services typically represent generic functionality, available `at a click'. Providers may therefore, for reasons of reputation or responsibility, 
seek to ensure that the AIaaS services they offer are being used by customers for `appropriate' purposes. 

This paper introduces and explores the concept whereby AIaaS providers uncover situations of possible service misuse by their customers. 
Illustrated through topical examples, we consider the technical usage patterns that could signal situations warranting scrutiny, and raise some of the legal and technical challenges of monitoring for misuse. 
In all, by introducing this concept, we indicate a potential area for further inquiry from a range of perspectives. 
\end{abstract}

 \begin{CCSXML}
<ccs2012>
<concept>
<concept_id>10010147.10010178</concept_id>
<concept_desc>Computing methodologies~Artificial intelligence</concept_desc>
<concept_significance>500</concept_significance>
</concept>
<concept>
<concept_id>10010520.10010521.10010537.10003100</concept_id>
<concept_desc>Computer systems organization~Cloud computing</concept_desc>
<concept_significance>500</concept_significance>
</concept>
</ccs2012>
<ccs2012>
<concept>
<concept_id>10003456.10003462</concept_id>
<concept_desc>Social and professional topics~Computing / technology policy</concept_desc>
<concept_significance>500</concept_significance>
</concept>
</ccs2012>
\end{CCSXML}

\ccsdesc[500]{Computing methodologies~Artificial intelligence}
\ccsdesc[500]{Computer systems organization~Cloud computing}
\ccsdesc[500]{Social and professional topics~Computing / technology policy}

\keywords{artificial intelligence; machine learning; cloud computing; law; accountability; misuse; monitoring; audit; compliance; MLaaS; AIaaS}

\maketitle

\section{Introduction}
\label{s:introduction}

\begin{textblock*}{20cm}(1cm,1.5cm) % {block width} (coords) 
   	{\centering{\color{red}
	
	{\fontsize{10}{12}\selectfont Pre-print: To appear in AAAI/ACM Artificial Intelligence, Ethics \& Society 2020, DOI: https://doi.org/10.1145/3375627.3375873 \par}\vspace{1.5mm}
	{\fontsize{7}{6}\selectfont Citation info: Seyyed Ahmad Javadi, Richard Cloete, Jennifer Cobbe, Michelle Seng Ah Lee and Jatinder Singh. `Monitoring Misuse for Accountable `Artificial Intelligence as a Service'''. \\In \textit{Proceedings of the 2020 AAAI/ACM Conference on AI, Ethics, and Society (AIES `20)}, ACM, New York, NY, USA, 2020.}
\par}	
}
\end{textblock*}

Artificial Intelligence (AI) has recently seen a surge of interest. It is touted to influence many aspects of modern society, including in areas such as health, agriculture, finance, transport, manufacturing, retail, education, science, government and public services, to name but a few. 

At the same time, AI is coming under increasing scrutiny \cite{Cobbe_2019,decprov}.
As the discussions of `algorithmic accountability', `AI regulation' and `ethical AI'  make clear, there are many situations in which the use of AI will be inappropriate, controversial and unlawful~\cite{helbing2019societal}.

We increasingly see providers offering \textit{`AI as a Service'} (AIaaS). 
In essence, AIaaS aims at providing the ML (machine learning)-driven building blocks to support customer applications~\cite{Parsaeefard_Tabrizian_Leon-Garcia_2019}.
It includes service providers offering access to a range of pre-built models and related services, whereby customers can send to the service particular inputs and receive back the results of a ML process, e.g. predictions, classifications, etc.
Offerings tend to be fairly generic\footnote{Though tailoring is possible, and some offer consulting services for specific needs.}---examples including object detection, text to voice synthesis, facial recognition, and text translation (see \S\ref{ss:types-of-services})---and therefore can be attractive to a range of customers, and suitable for driving applications across a number of scenarios and sectors. For instance, the same object recognition service used by one customer for warehousing might be used by another to support video surveillance.

Interesting considerations are raised by AIaaS, given the services make available sophisticated AI capabilities, often as `turnkey' (on demand, with a few clicks), to potentially anyone. 
That is, \textit{there is much scope for AIaaS services to be used for controversial and problematic purposes.}
As a provocative example, facial recognition is an AIaaS service offered by several providers, and intuitively has huge potential for misuse and abuse~\cite{EU_AFR_FR}.
In line with this, we have seen providers setting out principles of use~\cite{microsoft2019facial}, being active in related public discourse~\cite{google2019facial}, and implementing simple technical barriers to limit misuse (see \S\ref{s:overview}).
However even seemingly benign services also have the potential to drive controversial applications, e.g. generic object recognition services can assist military operations~\cite{whittaker_2018}.

This paper provides an initial exploration into monitoring as a means for facilitating greater oversight and governance of the use of AIaaS.
Providers appear to have an interest in such, be it
for reasons of reputation, e.g. to avoid the backlash where an unsavoury application could be labelled as \textit{`Powered by [OrgX]'} or \textit{`[ProviderY] Inside'}\texttrademark; or more generally, to help ensure that customer usage accords with their terms of service, principles, or requirements. 
Our concept of monitoring service usage might also work to inform future legal or regulatory regimes, which may demand more accountability from those providing access to AI\slash ML services. 

Specifically, we explore how AIaaS usage patterns could indicate situations warranting attention, and highlight challenges and opportunities for future work in the legal and technical spaces.

 \section{AIaaS overview}
\label{s:overview}

AI is a broad term. In the context of AIaaS, it typically refers to the use of machine learning (ML) (though we use `AI' and `AIaaS' to be consistent with the terminology used by providers in this space). ML works to uncover patterns in data, to build and refine representative \textit{models} of that data ~\cite{Qiu-Feng-2016}. 
These models can be used to make classifications, predictions, and so forth.

There is significant interest in using ML to underpin a range of applications.
However, undertaking ML, i.e. building models, can be challenging \cite{Zhou_Pan_Wang_Vasilakos_2017}.
Organisations may have issues regarding access to data -- given that models are built on data, one requires access to sufficient volumes of data to be representative of the problem space, and to enable model training and testing \cite{DeanJared2014Bddm}.
There are also issues of access to resources. Training models can be computationally expensive \cite{DeanJared2014Bddm}, so access to sufficient compute is a concern; as is ML expertise, which is said to be in short supply \cite{forbes}.

Recognising an opportunity, we increasingly see providers offering {AI as a Service} (AIaaS). 
The motivation for AIaaS is that many organisations will seek to leverage AI within their applications, but may lack the data, resources, capabilities, or time to undertake and maintain everything in-house. Indeed, this is similar to what drives the uptake of cloud infrastructure more generally. 
Unsurprisingly, we see that the prominent AIaaS providers are the tech giants, given their access to data, infrastructure and expertise.

Broadly there are two categories of AIaaS, those that (i) provide the infrastructure to support customers in undertaking their own machine learning; and those that (ii) provide access to a range of pre-built models and related services, whereby customers can send to the service particular inputs and receive back results, e.g. predictions, classifications, etc. 
Here we focus on the latter, using `AIaaS' to refer to such (though some of the aspects we raise may also be relevant to the more model building-oriented ML services).

In this way, by offering access to models, AIaaS essentially provides `application building blocks' which enables customers (tenants) to more easily integrate AI capabilities into the systems they build and run. 
One can expect AIaaS to grow in prominence. This is not only due to the increasing interest in AI, but also because customers benefit through lower barriers to entry: on-demand and pay-per-use reduces overheads in terms of engineering effort, cost and time-to-market, and where issues of maintenance and scalability are `outsourced' to providers.

\subsection{Types of services}
\label{ss:types-of-services}

 AIaaS entails service providers offering their customers (tenants) access to  models via APIs (Application Programming Interfaces -- these facilitate the interaction between systems).
 
 Typically, AIaaS offerings tend towards more generic tasks.
We observe that several AIaaS providers~\cite{microsoft-ai-service,google-ai-service}, group their offerings into four main categories:

\fakeheader{Language} services include text analytics (e.g. sentiment analysis), translation (e.g. automated text translation like Google Translate, language detection, etc.), language understanding and knowledge base creation (e.g. from collections of questions and answers).
    
\fakeheader{Analytics} services comprise capabilities for the analysis of data for various purposes, such as product recommendations (e.g. to deliver personalised ads), anomaly detection (e.g. for detecting when data behaviour changes), knowledge inference (e.g. to make predictions or forecasts based on your data) and content moderation (e.g. to detect offensive or unwanted images).

\fakeheader{Speech} services primarily comprise features such as text-to-speech, speech-to-text, speaker recognition (i.e. identifying individuals based on patterns in their speech), and so forth.

\fakeheader{Vision} services enable the analysing of images and videos in order to find and identify objects, text, and labels, just to name a few. 
Relevant to our later discussion, there are also offerings relating to faces: Microsoft's Azure Face~\cite{microsoft-ai-service} offers various APIs including face verification, face detection, emotion recognition, face identification, similar face search, celebrity recognition; Amazon Rekognition offers a similar range of face-oriented services~\cite{amazon-ai-services}.

\subsection{Accessing AIaaS}
\label{ss:accessing-aiaas}

\begin{figure}[htb]
\centering
\includegraphics[width=1.0\columnwidth]{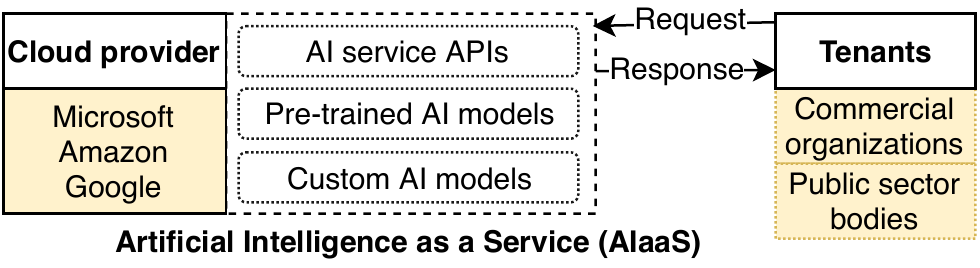} 
\caption{A simplified illustration of AIaaS in which cloud providers make AI technologies accessible to their customers (tenants).}
\label{fig:AIaaS-big-pic}
\end{figure}

To integrate AI into their applications, AIaaS customers setup and configure the services with possible customisations (which can entail model training). This is usually done through a specialised web interface supplied by the provider. Usage of the AIaaS service entails sending requests (e.g. `find all faces in this image') to the provider's API. The provider receives a request, and after conducting the required authentication and authorisation checks, processes the request given and returns the response (see Fig.~\ref{fig:AIaaS-big-pic}). 

\subsection{Transaction-oriented pricing}
\label{ss:transaction-oriented-pricing}

We explored the pricing model of three major providers, namely Amazon, Microsoft and Google, focusing on their vision-related AIaaS offerings.
Despite subtle differences, the pricing structures are largely similar across the providers.
In general, the API calls made can request one or more AI services (e.g., label detection, face recognition, object detection, etc.) to be applied to their input images. Azure's vision pricing model~\cite{microsoft-ai-service} refers to each service call as a \textit{transaction}. Several AIaaS service requests can be grouped into a single API call, but the tenant will be charged according to the total number of services requested. The Amazon Rekognition Image API pricing model for faces~\cite{amazon-ai-services} charges per image analysed and `for each set of facial feature vectors you store'. Google charges per image~\cite{google-ai-service}, where `each feature applied to an image is a billable {unit}', again meaning that the cost incurred per image will vary depending on the number of services applied to that image.

In short, all the major providers essentially bill in line with usage, relating to API calls (requests\slash responses).
This is relevant as such billing methods naturally entail some monitoring, if only for accountancy purposes. 
As such, 
billing may provide a starting point for monitoring for misuse, 
as it implies providers already employ some means for 
 monitoring transactions, and that some data is available regarding the transactions themselves.

% further elaborate on ways these services may be misused \noterc{Seems strange to next talk about incentives, given we've just indicated that audit logs captured from billing may be indicative of misuse.}\notejs{DOn't talk about misuse yet -- this is a background. Just comment the last bit out, we can bring it up latert}
 
\section{Potential for misuse}
\label{s:motivation}

The potential for the misuse of AI in general has been the subject of much attention.
There are a number of examples of inappropriate uses of machine learning. One example is the infamous attempt to use deep learning to predict a person's sexual orientation from a photograph, which was widely criticised for many reasons, including that such applications would have much potential for misuse in terms of supporting persecution and hate crime \cite{miller2018searching}. 
In the facial recognition context, there has been much debate regarding the acceptability of such technologies, leading some governments  to institute a ban on their use by the public sector~\cite{conger2019san}. The recent media coverage of police use of facial recognition on protesters in Hong Kong \cite{doffman2019hong-kong} has sparked fears of persecution. The installation of facial recognition systems in London raised privacy concerns~\cite{sabbagh2019facial}.

Given the discussion around ethical AI, algorithmic accountability, and so forth, technology companies have joined efforts in self-regulation \cite{rossi2019building}, and defining principles regarding AI.
For instance, the Partnership on AI, founded by Amazon, Facebook, Google, DeepMind, Microsoft, and IBM---who incidentally are the prime AIaaS providers---lists six core tenets of AI: 1) safety-critical, 2) fair, transparent, and accountable, 3) labor and economy, 4) collaboration with people, 5) social and societal influences, and 6) AI and social good \cite{rossi2019building}. Many have their own principles of AI ethics; Google's states that AI should protect privacy and be socially beneficial, fair, safe, and accountable to people, while IBM's focuses on trust and transparency \cite{rossi2019building}.

Despite this attention on the potential risks of AI, there has been considerably less focus on AI when offered as a service, and the potential for  cloud-based AI services to be misused by tenants.

We argue that this warrants consideration. 
 We have seen that seemingly innocuous technology can quickly   be re-purposed as the building blocks for nefarious applications. `Deepfakes', the use of AI to superimpose another person's face in a video, can be easily built using Google's open-sourced software with readily accessible tutorials, raising risks that it will be used to spread misinformation, threaten national security, or create pornography depicting non-consenting individuals~\cite{harris2018deepfakes}.

AIaaS presents a challenge, given that it deliberately aims to \textit{make widely accessible sophisticated AI that is capable of underpinning a broad range of applications}.
This `out-of-the-box' access to such functionality makes it likely that AIaaS will be used, maliciously or otherwise, by some  customers for controversial purposes.

\subsection{Monitoring AIaaS}

There are various reasons why providers might become active in monitoring and auditing how their AIaaS offerings are being used.
Monitoring and audit supports providers in taking a proactive role, by indicating situations that require attention and enabling the mitigation of negative outcomes. 

A key driver is \textit{reputation}.
Given the potential concerns regarding the use of AI for particular purposes, a controversial application being branded as \textit{`Powered by [Provider\_X]'} could lead to public backlash and undesirable attention. 
This is particularly the case for AIaaS, given that such services provide access to fairly generic models that could be used for a wide variety of purposes.

Providers also typically define in their terms of services particular responsibilities and obligations of use. 
Providers may therefore be encouraged to employ some monitoring as a means for indicating situations where breaches may be taking place.
Moreover, given that much value lies within the models being accessed, providers may seek to prevent data being leaked or aspects of the model being `stolen', e.g.  through model inversion attacks~\cite{tramer2016stealing}.

Indeed, there is evidence of AIaaS providers beginning to consider such issues, though so far this has tended only to concern specific, more obviously `risky' offerings -- particularly facial recognition.
For example, Google does not provide facial recognition as a service due to its potential for abuse~\cite{kent2018gooleai},
while Microsoft has been involved in the broader public discussion around such issues~\cite{smith2018facerecognition}
and denied police access to its facial recognition technology~\cite{vincent2019microsoft}. 
In terms of technical measures, Microsoft limits the request rate to their Face API, while Amazon prevents more than 100 faces from being detected in single image~\cite{amazon-ai-services},
both of which may work to mitigate misuse.
That said, more general frameworks regarding the appropriate uses of AIaaS, and how these would be monitored, so far have had little discussion. 

Conversely, providers may be hesitant to become more involved in monitoring the use of their services.
First, their customers (tenants) may not accept their activity being monitored, perhaps due to, for example, the nature or commercial sensitivity of their undertakings. Indeed, trust regarding cloud providers is a long-standing issue~\cite{siani}. 
Further, monitoring could result in an increased liability exposure for the provider (see \S5.1).

However,  there is also an argument that AIaaS providers \textit{should} play a greater role in monitoring their services. In addition to the points just mentioned, this is because providers (and indeed, their customers) profit from the offering of low-cost access to powerful ML services for use at scale.  
Moreover, while the misuse of cloud infrastructure services may be a general concern~\cite{lindemann2015towards,hashizume2013three}, AIaaS warrants particular consideration, not only due to the potential for AI misuse, but also because such services mean the provider plays a more direct role in enabling application functionality. As such, one could envisage providers incentivised to take a more proactive role, be it through their own volition, given the demands for greater tech-accountability, or as a result of the evolving regulatory landscape. 
We discuss some legal aspects in greater depth in \S\ref{ss:legal-challenges}, though reiterate that the purpose of this paper is to introduce the concept and the potential of AIaaS monitoring and audit, in order to provide the groundwork for future discussion and research.

% \noterc{perhaps add some text about how e.g. MS limits rates, Google doesn't offer the service.... these are examples that companies have taken steps or at least thought about the potential implications of providing such services.}
 \section{Monitoring for AI\lowercase{aa}S misuse } \label{s:miuse-detection}

\begin{table*}[!t]
\caption{Possible information sources to support the auditing of AIaaS usage.}
\centering
\begin{tabular}{|m{3cm}|m{2cm}|m{10.65cm}|}
\hline
\multicolumn{1}{|c|}{\textit{Category}} & \multicolumn{1}{c|}{\textit{Sub-category}} & \multicolumn{1}{c|}{\textit{Examples}}\\\hline
\multirow{3}{*}{Transaction raw data} & Metadata &  
Start time, client IP address, size of input parameters, resource usage, \ldots \\\cline{2-3}
 & Request & Input parameters, such as images, videos, audio files, \ldots \\\cline{2-3}
 & Response & Information returned to the clients during a transaction (e.g., detected faces)\\\hline 
\multicolumn{2}{|l|}{Request processing by-products} & Generated information (e.g., face encoding) as part of the request processing 
 \\\hline
\multicolumn{2}{|l|}{Derived data} & Data structures or ML models obtained from the raw and by-product data\\\hline 
\end{tabular}
\label{tabale:possible-aduit}
\end{table*}

There are several ways that providers could employ tighter governance regimes in AIaaS contexts. This could include, for instance, changing how services are offered, such as requiring initial consulting and application vetting rather than `turnkey' service access (as some do), requiring specifications of use inline with well-defined and context-specific terms of service provisions, and so forth. 

However, given the current and well-established model for cloud (and AIaaS) procurement---which is `turnkey', on-demand, through automatic, tech-driven processes---we consider how providers might technically discover that their AIaaS services are being used inappropriately. 
Our focus is in line with the current way AIaaS operates, where providers are fairly `hands-off' with regards to how their services are procured and integrated by customers. 

In essence, our concept involves exploring how \textit{monitoring and analysing customer usage of AIaaS services may help providers uncover situations warranting review or further investigation}. 
We emphasise that such an approach, which involves examining patterns of use,  transactions, etc., will often only be indicative of possible misuse rather than absolute, yet remains useful in highlighting potential problems as they occur and in enabling responses. This offers much potential compared to the alternative of `running blind'.

\begin{figure}[htb]
\centering
\includegraphics[width=1.0\columnwidth]{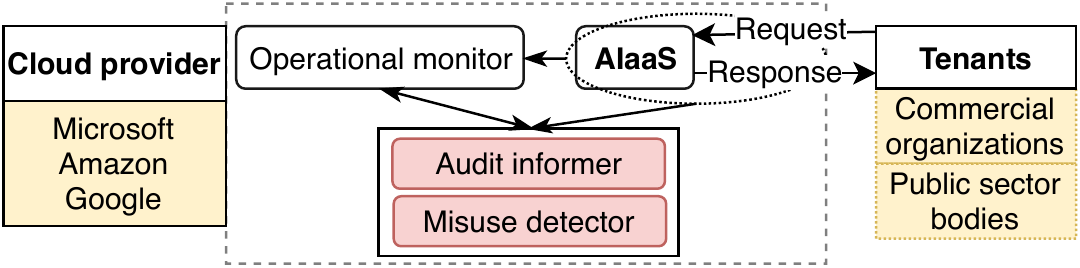}
\caption{A conceptual AIaaS misuse detection architecture.}
\label{fig:detection-generic-view}
\end{figure}

Fig.~\ref{fig:detection-generic-view} presents a high-level conceptual representation of a generic AIaaS misuse detection architecture. 
 The concept illustrates a component called \textit{Audit Informer}, which obtains and derives the relevant data to drive the analysis and detection of potential situations of concern. Some of this data might relate to that already collected by providers (via the \textit{Operational Monitor}) in order to, for example, enable billing (which is transaction oriented, see \S\ref{ss:transaction-oriented-pricing}), managing performance and other resources, etc. The \textit{Misuse Detector} component involves analysing and processing the audit information as a means for identifying and alerting of potential AIaaS misuse.

\subsection{Audit information}
\label{ss:monitoring-level}

Detecting AIaaS misuse requires information. There may be a range of `signals' available to a cloud provider.
Table~\ref{tabale:possible-aduit} presents some categories of information that may be relevant, which include: 
\begin{itemize}
    \item \textit{Transaction raw data:} Includes three main sub-categories, namely request metadata (e.g., timestamp, size, account credentials), request specifics (e.g., input parameters such as images, audio clips, etc.), and response specifics (e.g., outputs from the model such as predictions or classifications). Note that some of this may already be collected by providers (Operation Monitor), e.g. to support billing. 
    \item \textit{Request processing by-products:} Includes data generated in serving the request, which can contain, for example, any input data pre-processing, internal processing pipelines, collation of results from multiple services, intermediary feature vectors, etc.
    \item \textit{Derived data:} Concerns data processed or derived from transaction logs, request by products, etc, to assist detection and analysis. This could include, for example, creating structured log files, bloom filters, models and representations of typical usage (encoding usage patterns, inputs, output results), etc.  
    
\end{itemize}

\subsection{Uncovering misuse}
\label{ss:misuse-detection}

\begin{table*}[h]
\caption{Some examples of some AIaaS misuse indicators for object and facial recognition services.}
\centering
\begin{tabular}{|m{10.5cm}|m{5.5cm}|}
\hline
\multicolumn{1}{|c|}{\textit{Technical indicator for vision and face services} } & \multicolumn{1}{c|}{\textit{Potential implications}} \\\hline 
High request rate for face detection & 
Population surveillance \\\hline 
Large number of faces in an image/video & Population surveillance \\\hline 
Large number of different faces are analysed & 
Population surveillance \\\hline 
Large number of identification attempts for particular individual(s)
& Privacy threats to an individual \\\hline 
Detection of `black-listed' objects
& Controversial application  \\\hline 
\end{tabular}
\label{tabale:indicators}
\end{table*}

We now discuss detecting potential cases of misuse. The concept entails \textit{misuse indicators}, which reflect certain criteria of tenant behaviour that may warrant some, or indeed, further attention.
The implementation of an indicator entails an analysis of the relevant audit information sources to determine whether the particular criteria has been met.
In practice, there are many possible types and forms of indicators, some generic and some context-specific.
In this section we present the general concept to offer a way forward,
using facial recognition as an illustrative example~\cite{EU_AFR_FR}.
 
\subsubsection{Specific misuse indicators}
There will be situations in which providers will have some knowledge or foresight as to the particular risks of a particular service, or the forms that misuse might take. 
For instance, in facial recognition contexts, some concerns relate to population surveillance, personal privacy, and use by state services~\cite{zeng2019responsible,brundage2018malicious}. 

Table~\ref{tabale:indicators} presents some specific examples of misuse indicators for vision and face-based AIaaS services and their category of risk. We now elaborate these (with reference to Fig.~\ref{fig:detection-generic-view} and Table~\ref{tabale:possible-aduit}):
\begin{itemize}
    \item\textit{High request rate for face services}: If the tenant's request rate for face services is high, it possibly indicates facial recognition deployment at-scale (e.g., population surveillance). Indeed, we see a rate limiting approach already employed by Microsoft~\cite{microsoft-ai-service}. Request rate patterns are easily derived from the transaction metadata already captured by platforms. 
    \item \textit{Large number of faces in an image\slash video}: Also concerned with surveillance, this indicator targets whether tenants appear to be analysing crowds by looking for many faces within the same input. We note Amazon AWS limits the number of faces detected in any image~\cite{amazon-ai-services}, though it is unclear if this is due to surveillance concerns.
    \item \textit{Large number of different faces are analysed }: A consistent use of a service to uncover different faces \textit{over time} could also indicate a population surveillance scenario. 
    This indicator could include considering aspects such as the frequency and volume of faces matched (i.e. numbers regarding input\slash output faces), to more sophisticated (and potentially more intrusive and legally-challenging) approaches, which might involve, for example, recording and analysing the raw input images or other details of the faces involved.     \item \textit{Large number of identification attempts for particular individual(s)}:\footnote{Identification in this context is described as a one-to-many comparison~\cite{EU_AFR_FR}, i.e. `looking for' individual(s).} This concerns a tenant searching for a similar face across a large number of images coming from different contexts, suggesting the targeting and tracking of particular individual(s). 
    Such an analysis might involve keeping records of the inputs\slash outputs to particular services, the results of processing, etc.
        \item \textit{Detection of `black-listed' objects}: This indicator looks for certain categories of items that might generally be of a type indicating a problematic application. For example, repeated detection of placards, or `violence' could indicate that the technology is being used to screen protests. 
    Moreover, this example illustrates the relevance of monitoring \textit{across} different AIaaS services; e.g. detecting blacklisted objects and faces from the same inputs, such as placards (object detection) and multiple faces (facial recognition), may be revealing.   
    
\end{itemize}

\subsubsection{Misuse discovery through patterns of behaviour}
\label{anom}
There will also be situations where a specific risk might not have been foreseen, but where the usage pattern itself might indicate a need for further scrutiny. 
This entails building profiles of AIaaS usage for and across particular services. Analysis might involve using, for example, anomaly detection methods~\cite{chandola2009anomaly} to indicate potentially problematic usage patterns. This is similar to how such methods are used in the security domain, e.g. for intrusion detection, where changes in network traffic can indicate that a system has been compromised~\cite{ahmed2016survey}.

This idea could useful by indicating 
where the behaviour of a customer deviates from their normal usage pattern. For instance, if a customer transaction begins using a face recognition API far more extensively than previously, or integrates face detection with other services having not previously done so, it might warrant further investigation. 
Similarly, one can
 compare the usage profiles across customers. Assuming most customers (as organisations with responsibilities) are behaving appropriately, detecting a particular usage pattern that contrasts to that of others may  warrant attention.

A further example concerns patterns indicating a model inversion attack~\cite{fredrikson2015model}, whereby tenants may be attempting to reverse-engineer the classification-criteria or training data from the model itself. Given the value of AIaaS is in the model, and the potential legal implications should models encode personal data (see \cite{inversion}), this might be of particular concern for providers. Such attacks may require coordination from across a range of accounts, and there appears space for exploring the use of ML\slash anomaly detection towards detecting such situations. 

 \section{Practical challenges \& opportunities}
\label{s:practical-challenge}

So far we have introduced the concept of AIaaS monitoring as a means for indicating possible misuse. This section discusses some relevant legal and technical considerations,  highlighting some interesting practical challenges and opportunities for future work.

\subsection{Legal}
\label{ss:legal-challenges}

AIaaS providers monitoring customers' use of their services and models may have legal consequences. We now explore these in relation to data protection law and intermediary liability, noting this is an area requiring further consideration. As such, rather than undertaking an in-depth analysis of the various legal issues and questions potentially arising from monitoring AIaaS, we highlight some considerations with a view to exploring these questions in greater detail in subsequent work. Note that, as well as the issues discussed below, further considerations are likely to arise in relation to the enforcement of contractual provisions and terms of service applying between AIaaS providers and their customers.

\subsubsection{Data protection}

Under the EU's General Data Protection Regulation (GDPR)~\cite{gdpr} (and similar legislation in other jurisdictions), some of the audit information described above may be classed as \textit{personal data} (i.e. any information relating to an identified or identifiable natural person [GDPR Art 4.1]). Depending on the specifics, this would likely include much information that we have categorised in Table 1 as `transaction raw data', such as metadata (IP address), input parameters, and the information returned to clients, and some kinds of processing by-products. Note, however, that the operational metrics described earlier and some derived data are unlikely to be personal data, should they entail, for instance, aggregated or more general information about customers' use of services (e.g., regarding transactions, rates of service use, etc).

GDPR establishes a framework governing the \textit{processing} of personal data (essentially any use or operation on personal data [GDPR Art 4.2]), with strict compliance requirements and potentially serious penalties for non-compliance. AIaaS providers who process personal data in monitoring customers' use of their services, as in some of the monitoring methods this paper proposes, are likely to be data controllers [GDPR Art 4.7] for that monitoring -- as they determine its means (how they are doing it) and its purposes (why they are doing it). 
This means they would then be subject to the GDPR's extensive compliance obligations [GDPR Art 5.2]. 

Providers may in practice be unable to determine which of the 
data that they are processing is personal data without first processing it. As a result, providers might best, in processing such kinds of audit information, take a precautionary approach that treats all of the data that has the potential to be personal as personal data. 
Moreover, GDPR recognises certain `special categories' of personal data that are particularly sensitive [GDPR Art 9]. 
Given the inability to determine which data falls into one of these categories without processing it, cloud providers may need to treat large swathes of data for monitoring as if it was special category. 

All personal data processing requires a basis in law [GDPR Arts 5, 6, 9]. The available bases for processing special category data are restricted [GDPR Art 9], and it's likely that only \textit{explicit} consent of data subjects (individuals to whom personal data relates) would be available in these circumstances. It is not obvious how AIaaS providers could identify the data subjects to seek or obtain their explicit consent to the processing for monitoring without first processing the data in question. In circumstances where personal data is likely to be involved, cloud providers monitoring tenants' use of models would therefore face a dilemma. They will not in many cases be able to monitor without the explicit consent of any data subjects whose special category data will be processed in doing so. Without processing the data, they would be unlikely to be able to identify data subjects in order to obtain their explicit consent.

\subsubsection{Intermediary liability}

Many jurisdictions provide online platforms with protection from liability for any hosted content that might be illegal or unlawful in some way. In different jurisdictions, this protection may be given on a blanket basis, or may be qualified in some way or subject to certain conditions. In the EU, this is provided for in the E-Commerce Directive (ECD) \cite{ecomdirective}. Protection from liability is available to service providers who store information on behalf of users of their services by passive, technical means, so long as the provider does not have actual knowledge of any illegality [ECD Art 14, recital 42]. If the provider acquires such knowledge, they are obliged to remove the illegal information or activity expeditiously [ECD Art 14].

It is uncertain whether cloud providers offering AIaaS could avail of liability protections for hosting, due to the nature of the active role they play in providing that service. The current jurisprudence of the European Court of Justice indicates that, in some circumstances, service providers who take an active role may go beyond activities for which protection from liability is provided \cite{googleFrance}\cite{eBayCJEU}. However, the Court's case law is currently unclear on precisely which kinds of activities would involve a provider taking such a role (although note that, as of December 2019, the Court is considering a case that may provide an opportunity the clarify the law in this area \cite{YouTubeCJEU}). 

Here, AIaaS providers are not simply offering a hosting facility by purely passive, technical means, but rather, offer access to sophisticated models that classify, predict, make decisions, etc. Providers may therefore have some liability for their use. However, if they \textit{can} ordinarily avail of that protection, actively monitoring tenants' activities could take them outside of that protection.
In order to avoid legal repercussions, they would be obliged to expeditiously remove or prevent access to any content or activity that their monitoring determined to be illegal. Cloud providers might therefore prefer to `play dumb' in order to avoid triggering any obligations and legal risks under the Directive.
This is an interesting area requiring further investigation.

\subsubsection{Future}

Note, however, that in some jurisdictions, including the EU, there are proposals for reforming the law around the responsibilities of online platforms.
 Several of these propose to place greater responsibilities on platforms around the information they host or process (e.g.~\cite{onlineharms,copyright}). This may put providers under obligations to monitor use of their models (which, given that they are providing access to powerful probabilistic models at scale and with a low barrier to entry, may not be an unreasonable development). If monitoring became explicitly required by law, that could potentially also allow them to circumvent the problems with obtaining explicit consent from data subjects. This is because GDPR provides a legal basis for processing special category data where doing so is necessary for reasons of substantial public interest and is undertaken on the basis of EU or Member State law that provides for sufficient safeguards for data subject rights [GDPR Art 9].  \subsection{Technical}
\label{ss:technical}
Monitoring for AIaaS misuse by platforms also raises a number interesting technical challenges and opportunities for research.

A key area of research concerns the technical mechanisms to support oversight, which includes (i) technical monitoring architectures, and (ii) the analysis and development of misuse indicators.  
Some approaches may aim to operate generally, while others may be more context\slash application-specific. 
Work on intrusion detection and threat monitoring in networked contexts (\S\ref{anom}), as well as work on provenance~\cite{decprov}, appear relevant as possible starting points. 
Testing and ensuring that the indicators are meaningful and accurate is also an area for  consideration. 

Another aspect relates to performance. Service providers naturally seek to optimise the performance and reduce the resource requirements of the services that they offer. At the same time, monitoring entails overhead (e.g. processing, storage, etc.). As such, there are opportunities for research into quantifying the overheads and other impacts of various monitoring and detection processes. Note again that some aspects, such as request metadata, are likely already to be captured by providers, such as to assist with billing (\S\ref{ss:transaction-oriented-pricing}), and so there may be ways to develop monitoring methods that leverage these to avoid adding substantial extra overheads.

Further, and in line with the above discussion, there may well be legal issues and liability implications from providers undertaking monitoring, and more generally, being more actively involved in policing the use of their services. The specifics of any potential increase in liability exposure will certainly depend on the circumstances. 
Such concerns also represent opportunities for technical research; examples include devising indicators and methods for monitoring that are more privacy and data-protection `friendly'.

% An important technical challenge is that AIaaS misuse detection imposes development, storage and computing, and maintenance overhead to cloud providers. Therefore, a comprehensive cost-sensitivity analysis is our future research work. In this context, cost refers to how much manpower, and storage and computing resources are used for AIaaS misuse detection. On the other hand, sensitivity refers to the monitoring level and how the misuse detection component parameters are set (please see figure~\ref{fig:detection-generic-view}). If we can map the cost and sensitivity to concrete values and draw the former based on the latter, we would like to see their relationship as an example is shown in figure~\ref{fig:cost-sensitivity}. Accordingly, we want to study how the cost increases as we monitor more information about AIaaS transactions and conduct heavier analysis or be more sensitive to any suspicious cloud tenants' behaviour.
  \section{Conclusion}
AIaaS makes accessible a wide range of sophisticated and scalable AI services for customers to integrate into their applications. These services act as application `building blocks', which are potentially accessible by anyone. `Turnkey' access to such services reduces the direct interactions with the provider, and so too the opportunities for oversight and intervention. 
It is therefore foreseeable that there will be situations in which AIaaS is used for controversial purposes, be it intentionally or otherwise. 

While issues surrounding AI ethics, regulation and responsibility are currently the subject of much discussion, the potential for AIaaS misuse has had comparatively little attention. This is important, as given the complexity, data and skills required for ML undertakings, it is likely that AIaaS will become a dominant part of the technical infrastructure underpinning AI-driven applications. 
There appears a role for  AIaaS providers to be more active in governing the use of their offerings, be it for reasons of reputation or responsibility -- not least given the growing demands for greater tech-accountability. 
In this paper, we used illustrative examples to highlight related issues, providing an initial conceptual overview of how providers might work towards ensuring more responsible use of their services. We also indicated the potential challenges and opportunities for moving forward, both from a legal and technical perspective. 

In all, by presenting this concept, we seek to draw attention to AIaaS and its usage as an area warranting further  consideration.

\balance

\section*{Acknowledgments}
We acknowledge the financial support of the Engineering \& Physical Sciences Research Council, University of Cambridge, Microsoft via the Microsoft Cloud Computing Research Centre, and Aviva.

\bibliographystyle{ACM-Reference-Format}

\begin{thebibliography}{41}


\ifx \showCODEN    \undefined \def \showCODEN     #1{\unskip}     \fi
\ifx \showDOI      \undefined \def \showDOI       #1{#1}\fi
\ifx \showISBNx    \undefined \def \showISBNx     #1{\unskip}     \fi
\ifx \showISBNxiii \undefined \def \showISBNxiii  #1{\unskip}     \fi
\ifx \showISSN     \undefined \def \showISSN      #1{\unskip}     \fi
\ifx \showLCCN     \undefined \def \showLCCN      #1{\unskip}     \fi
\ifx \shownote     \undefined \def \shownote      #1{#1}          \fi
\ifx \showarticletitle \undefined \def \showarticletitle #1{#1}   \fi
\ifx \showURL      \undefined \def \showURL       {\relax}        \fi
\providecommand\bibfield[2]{#2}
\providecommand\bibinfo[2]{#2}
\providecommand\natexlab[1]{#1}
\providecommand\showeprint[2][]{arXiv:#2}

\bibitem[\protect\citeauthoryear{Ahmed, Mahmood, and Hu}{Ahmed
  et~al\mbox{.}}{2016}]        {ahmed2016survey}
\bibfield{author}{\bibinfo{person}{Mohiuddin Ahmed},
  \bibinfo{person}{Abdun~Naser Mahmood}, {and} \bibinfo{person}{Jiankun Hu}.}
  \bibinfo{year}{2016}\natexlab{}.
\newblock \showarticletitle{A Survey of Network Anomaly Detection Techniques}.
\newblock \bibinfo{journal}{\emph{Journal of Network and Computer
  Applications}}  \bibinfo{volume}{60} (\bibinfo{year}{2016}),
  \bibinfo{pages}{19--31}.
\newblock


\bibitem[\protect\citeauthoryear{Amazon}{Amazon}{2019}]        {amazon-ai-services}
\bibfield{author}{\bibinfo{person}{Amazon}.} \bibinfo{year}{2019}\natexlab{}.
\newblock \bibinfo{booktitle}{\emph{{Build an AI-driven Application}}}.
\newblock
\urldef\tempurl\url{https://aws.amazon.com/machine-learning/ai-services/}
\showURL{\tempurl}


\bibitem[\protect\citeauthoryear{Brundage, Avin, Clark, Toner, Eckersley,
  Garfinkel, Dafoe, Scharre, Zeitzoff, Filar, et~al\mbox{.}}{Brundage
  et~al\mbox{.}}{2018}]        {brundage2018malicious}
\bibfield{author}{\bibinfo{person}{Miles Brundage}, \bibinfo{person}{Shahar
  Avin}, \bibinfo{person}{Jack Clark}, \bibinfo{person}{Helen Toner},
  \bibinfo{person}{Peter Eckersley}, \bibinfo{person}{Ben Garfinkel},
  \bibinfo{person}{Allan Dafoe}, \bibinfo{person}{Paul Scharre},
  \bibinfo{person}{Thomas Zeitzoff}, \bibinfo{person}{Bobby Filar},
  {et~al\mbox{.}}} \bibinfo{year}{2018}\natexlab{}.
\newblock \showarticletitle{{The Malicious Use of Artificial Intelligence:
  Forecasting, Prevention, and Mitigation}}.
\newblock \bibinfo{journal}{\emph{arXiv preprint arXiv:1802.07228}}
  (\bibinfo{year}{2018}).
\newblock


\bibitem[\protect\citeauthoryear{Chandola, Banerjee, and Kumar}{Chandola
  et~al\mbox{.}}{2009}]        {chandola2009anomaly}
\bibfield{author}{\bibinfo{person}{Varun Chandola}, \bibinfo{person}{Arindam
  Banerjee}, {and} \bibinfo{person}{Vipin Kumar}.}
  \bibinfo{year}{2009}\natexlab{}.
\newblock \showarticletitle{Anomaly Detection: A Survey}.
\newblock \bibinfo{journal}{\emph{ACM computing surveys (CSUR)}}
  \bibinfo{volume}{41}, \bibinfo{number}{3} (\bibinfo{year}{2009}),
  \bibinfo{pages}{15}.
\newblock


\bibitem[\protect\citeauthoryear{Cobbe}{Cobbe}{2019}]        {Cobbe_2019}
\bibfield{author}{\bibinfo{person}{J. Cobbe}.} \bibinfo{year}{2019}\natexlab{}.
\newblock \showarticletitle{Administrative Law and the Machines of Government:
  Judicial Review of Automated Public-Sector Decision-Making}.
\newblock \bibinfo{journal}{\emph{Legal Studies}} (\bibinfo{year}{2019}).
\newblock


\bibitem[\protect\citeauthoryear{Conger, Fausset, and Kovaleski}{Conger
  et~al\mbox{.}}{2019}]        {conger2019san}
\bibfield{author}{\bibinfo{person}{Kate Conger}, \bibinfo{person}{Richard
  Fausset}, {and} \bibinfo{person}{Serge~F Kovaleski}.}
  \bibinfo{year}{2019}\natexlab{}.
\newblock \bibinfo{booktitle}{\emph{{San Francisco Bans Facial Recognition
  Technology}}}.
\newblock
\urldef\tempurl\url{https://www.nytimes.com/2019/05/14/us/facial-recognition-ban-san-francisco.html}
\showURL{\tempurl}


\bibitem[\protect\citeauthoryear{Dan}{Dan}{2019}]        {sabbagh2019facial}
\bibfield{author}{\bibinfo{person}{Sabbagh Dan}.}
  \bibinfo{year}{2019}\natexlab{}.
\newblock \bibinfo{booktitle}{\emph{{Facial Recognition Technology Scrapped at
  King's Cross Site}}}.
\newblock
\urldef\tempurl\url{https://www.theguardian.com/technology/2019/sep/02/facial-recognition-technology-scrapped-at-kings-cross-development}
\showURL{\tempurl}


\bibitem[\protect\citeauthoryear{Dean}{Dean}{2014}]        {DeanJared2014Bddm}
\bibfield{author}{\bibinfo{person}{Jared Dean}.}
  \bibinfo{year}{2014}\natexlab{}.
\newblock \bibinfo{booktitle}{\emph{Big data, data mining, and machine learning
  value creation for business leaders and practitioners / Jared Dean.}}
\newblock \bibinfo{publisher}{Wiley}, \bibinfo{address}{Hoboken}.
\newblock
\showISBNx{9781118920695}


\bibitem[\protect\citeauthoryear{{Department for Digital, Culture, Media \&
  Sport}}{{Department for Digital, Culture, Media \& Sport}}{2019}]        {onlineharms}
\bibfield{author}{\bibinfo{person}{{Department for Digital, Culture, Media \&
  Sport}}.} \bibinfo{year}{2019}\natexlab{}.
\newblock \bibinfo{booktitle}{\emph{{Online Harms White Paper, CP 57}}}.
\newblock
\urldef\tempurl\url{https://assets.publishing.service.gov.uk/government/uploads/system/uploads/attachment_data/file/793360/Online_Harms_White_Paper.pdf}
\showURL{\tempurl}


\bibitem[\protect\citeauthoryear{for Fundamental~Rights}{for
  Fundamental~Rights}{2019}]        {EU_AFR_FR}
\bibfield{author}{\bibinfo{person}{European Union~Agency for
  Fundamental~Rights}.} \bibinfo{year}{2019}\natexlab{}.
\newblock \bibinfo{booktitle}{\emph{{Facial Recognition Technology: Fundamental
  Rights Considerations in the Context of Law Enforcement}}}.
\newblock
\urldef\tempurl\url{https://fra.europa.eu/en/publication/2019/facial-recognition}
\showURL{\tempurl}


\bibitem[\protect\citeauthoryear{Fredrikson, Jha, and Ristenpart}{Fredrikson
  et~al\mbox{.}}{2015}]        {fredrikson2015model}
\bibfield{author}{\bibinfo{person}{Matt Fredrikson}, \bibinfo{person}{Somesh
  Jha}, {and} \bibinfo{person}{Thomas Ristenpart}.}
  \bibinfo{year}{2015}\natexlab{}.
\newblock \showarticletitle{{Model Inversion Attacks that Exploit Confidence
  Information and Basic Countermeasures}}. In \bibinfo{booktitle}{\emph{Proc.
  ACM SIGSAC}}. ACM, \bibinfo{pages}{1322--1333}.
\newblock


\bibitem[\protect\citeauthoryear{Google}{Google}{2019a}]        {google-ai-service}
\bibfield{author}{\bibinfo{person}{Google}.} \bibinfo{year}{2019}\natexlab{a}.
\newblock \bibinfo{booktitle}{\emph{{AI and machine learning products}}}.
\newblock
\urldef\tempurl\url{https://cloud.google.com/products/ai/building-blocks/}
\showURL{\tempurl}


\bibitem[\protect\citeauthoryear{Google}{Google}{2019b}]        {google2019facial}
\bibfield{author}{\bibinfo{person}{Google}.} \bibinfo{year}{2019}\natexlab{b}.
\newblock \bibinfo{booktitle}{\emph{Our approach to facial recongition}}.
\newblock
\urldef\tempurl\url{https://ai.google/responsibilities/facial-recognition/}
\showURL{\tempurl}


\bibitem[\protect\citeauthoryear{Harris}{Harris}{2018}]        {harris2018deepfakes}
\bibfield{author}{\bibinfo{person}{Douglas Harris}.}
  \bibinfo{year}{2018}\natexlab{}.
\newblock \showarticletitle{Deepfakes: False Pornography Is Here and the Law
  Cannot Protect You}.
\newblock \bibinfo{journal}{\emph{Duke L. \& Tech. Rev.}}  \bibinfo{volume}{17}
  (\bibinfo{year}{2018}), \bibinfo{pages}{99}.
\newblock


\bibitem[\protect\citeauthoryear{Hashizume, Yoshioka, and Fernandez}{Hashizume
  et~al\mbox{.}}{2013}]        {hashizume2013three}
\bibfield{author}{\bibinfo{person}{Keiko Hashizume}, \bibinfo{person}{Nobukazu
  Yoshioka}, {and} \bibinfo{person}{Eduardo~B Fernandez}.}
  \bibinfo{year}{2013}\natexlab{}.
\newblock \showarticletitle{{Three Misuse Patterns for Cloud Computing}}.
\newblock In \bibinfo{booktitle}{\emph{Security Engineering for Cloud
  Computing: Approaches and Tools}}. \bibinfo{publisher}{IGI Global},
  \bibinfo{pages}{36--53}.
\newblock


\bibitem[\protect\citeauthoryear{Helbing}{Helbing}{2019}]        {helbing2019societal}
\bibfield{author}{\bibinfo{person}{Dirk Helbing}.}
  \bibinfo{year}{2019}\natexlab{}.
\newblock \showarticletitle{Societal, economic, ethical and legal challenges of
  the digital revolution: from big data to deep learning, artificial
  intelligence, and manipulative technologies}.
\newblock In \bibinfo{booktitle}{\emph{Towards Digital Enlightenment}}.
  \bibinfo{publisher}{Springer}, \bibinfo{pages}{47--72}.
\newblock


\bibitem[\protect\citeauthoryear{James}{James}{2019}]        {vincent2019microsoft}
\bibfield{author}{\bibinfo{person}{Vincent James}.}
  \bibinfo{year}{2019}\natexlab{}.
\newblock \bibinfo{booktitle}{\emph{{Microsoft Denied Police Facial Recognition
  Tech Over Human Rights Concerns}}}.
\newblock
\urldef\tempurl\url{https://www.theverge.com/2019/4/17/18411757/microsoft-facial-recognition-sales-refused-police-access}
\showURL{\tempurl}


\bibitem[\protect\citeauthoryear{Lindemann}{Lindemann}{2015}]        {lindemann2015towards}
\bibfield{author}{\bibinfo{person}{Jens Lindemann}.}
  \bibinfo{year}{2015}\natexlab{}.
\newblock \showarticletitle{{Towards Abuse Detection and Prevention in IaaS
  Cloud Computing}}. In \bibinfo{booktitle}{\emph{2015 10th International
  Conference on Availability, Reliability and Security}}. IEEE,
  \bibinfo{pages}{211--217}.
\newblock


\bibitem[\protect\citeauthoryear{Marr}{Marr}{2018}]        {forbes}
\bibfield{author}{\bibinfo{person}{Bernard Marr}.}
  \bibinfo{year}{2018}\natexlab{}.
\newblock \bibinfo{title}{The AI Skills Crisis And How To Close The Gap}.
\newblock
\newblock
\urldef\tempurl\url{https://www.forbes.com/sites/bernardmarr/2018/06/25/the-ai-skills-crisis-and-how-to-close-the-gap/}
\showURL{\tempurl}


\bibitem[\protect\citeauthoryear{Microsoft}{Microsoft}{2019a}]        {microsoft-ai-service}
\bibfield{author}{\bibinfo{person}{Microsoft}.}
  \bibinfo{year}{2019}\natexlab{a}.
\newblock \bibinfo{booktitle}{\emph{{Cognitive Services}}}.
\newblock
\urldef\tempurl\url{https://azure.microsoft.com/en-gb/services/cognitive-services}
\showURL{\tempurl}


\bibitem[\protect\citeauthoryear{Microsoft}{Microsoft}{2019b}]        {microsoft2019facial}
\bibfield{author}{\bibinfo{person}{Microsoft}.}
  \bibinfo{year}{2019}\natexlab{b}.
\newblock \bibinfo{booktitle}{\emph{Six Principles For Developing and Deploying
  Facial Recognition Technology}}.
\newblock
\urldef\tempurl\url{https://blogs.microsoft.com/wp-content/uploads/prod/sites/5/2018/12/MSFT-Principles-on-Facial-Recognition.pdf}
\showURL{\tempurl}


\bibitem[\protect\citeauthoryear{Miller}{Miller}{2018}]        {miller2018searching}
\bibfield{author}{\bibinfo{person}{Arianne~E Miller}.}
  \bibinfo{year}{2018}\natexlab{}.
\newblock \showarticletitle{{Searching for Gaydar: Blind Spots in the Study of
  Sexual Orientation Perception}}.
\newblock \bibinfo{journal}{\emph{Psychology \& Sexuality}}
  \bibinfo{volume}{9}, \bibinfo{number}{3} (\bibinfo{year}{2018}),
  \bibinfo{pages}{188--203}.
\newblock


\bibitem[\protect\citeauthoryear{of~Justice}{of~Justice}{[n.d.]}]        {YouTubeCJEU}
\bibfield{author}{\bibinfo{person}{European~Court of Justice}.}
  \bibinfo{year}{[n.d.]}\natexlab{}.
\newblock \bibinfo{title}{LF v YouTube (C-682/18)}.
\newblock
\newblock


\bibitem[\protect\citeauthoryear{of~Justice}{of~Justice}{2010}]        {googleFrance}
\bibfield{author}{\bibinfo{person}{European~Court of Justice}.}
  \bibinfo{year}{2010}\natexlab{}.
\newblock \bibinfo{title}{Google France SARL and Google Inc. v Louis Vuitton
  (C-236/08) ECLI:EU:C:2010:159}.
\newblock
\newblock


\bibitem[\protect\citeauthoryear{of~Justice}{of~Justice}{2011}]        {eBayCJEU}
\bibfield{author}{\bibinfo{person}{European~Court of Justice}.}
  \bibinfo{year}{2011}\natexlab{}.
\newblock \bibinfo{title}{L'Orèal SA and Others v eBay International AG and
  Others (C-324/09) ECLI:EU:C:2011:474}.
\newblock
\newblock


\bibitem[\protect\citeauthoryear{Parsaeefard, Tabrizian, and
  Leon-Garcia}{Parsaeefard et~al\mbox{.}}{2019}]        {Parsaeefard_Tabrizian_Leon-Garcia_2019}
\bibfield{author}{\bibinfo{person}{Saeedeh Parsaeefard}, \bibinfo{person}{Iman
  Tabrizian}, {and} \bibinfo{person}{Alberto Leon-Garcia}.}
  \bibinfo{year}{2019}\natexlab{}.
\newblock \bibinfo{title}{{Artificial Intelligence as a Services (AI-aaS) on
  Software-Defined Infrastructure}}.
\newblock
\newblock
\showeprint[arxiv]{cs.LG/1907.05505}


\bibitem[\protect\citeauthoryear{Pearson and Benameur}{Pearson and
  Benameur}{2010}]        {siani}
\bibfield{author}{\bibinfo{person}{Siani Pearson} {and}
  \bibinfo{person}{Azzedine Benameur}.} \bibinfo{year}{2010}\natexlab{}.
\newblock \showarticletitle{Privacy, Security and Trust Issues Arising from
  Cloud Computing}. In \bibinfo{booktitle}{\emph{Proceedings of the 2010 IEEE
  Second International Conference on Cloud Computing Technology and Science}}
  \emph{(\bibinfo{series}{CLOUDCOM '10})}. \bibinfo{publisher}{IEEE Computer
  Society}, \bibinfo{address}{Washington, DC, USA}, \bibinfo{pages}{693--702}.
\newblock
\showISBNx{978-0-7695-4302-4}


\bibitem[\protect\citeauthoryear{Qiu, Wu, Ding, Xu, and Feng}{Qiu
  et~al\mbox{.}}{2016}]        {Qiu-Feng-2016}
\bibfield{author}{\bibinfo{person}{Junfei Qiu}, \bibinfo{person}{Qihui Wu},
  \bibinfo{person}{Guoru Ding}, \bibinfo{person}{Yuhua Xu}, {and}
  \bibinfo{person}{Shuo Feng}.} \bibinfo{year}{2016}\natexlab{}.
\newblock \showarticletitle{{A Survey of Machine Learning for Big Data
  Processing}}.
\newblock \bibinfo{journal}{\emph{Eurasip Journal on Advances in Signal
  Processing}} \bibinfo{volume}{2016}, \bibinfo{number}{1}
  (\bibinfo{year}{2016}).
\newblock


\bibitem[\protect\citeauthoryear{Rossi}{Rossi}{2019}]        {rossi2019building}
\bibfield{author}{\bibinfo{person}{Francesca Rossi}.}
  \bibinfo{year}{2019}\natexlab{}.
\newblock \showarticletitle{Building Trust in Artificial Intelligence}.
\newblock \bibinfo{journal}{\emph{Journal of international affairs}}
  \bibinfo{volume}{72}, \bibinfo{number}{1} (\bibinfo{year}{2019}),
  \bibinfo{pages}{127--134}.
\newblock


\bibitem[\protect\citeauthoryear{{Singh}, {Cobbe}, and {Norval}}{{Singh}
  et~al\mbox{.}}{2019}]        {decprov}
\bibfield{author}{\bibinfo{person}{J. {Singh}}, \bibinfo{person}{J. {Cobbe}},
  {and} \bibinfo{person}{C. {Norval}}.} \bibinfo{year}{2019}\natexlab{}.
\newblock \showarticletitle{Decision Provenance: Harnessing Data Flow for
  Accountable Systems}.
\newblock \bibinfo{journal}{\emph{IEEE Access}}  \bibinfo{volume}{7}
  (\bibinfo{year}{2019}), \bibinfo{pages}{6562--6574}.
\newblock
\showISSN{2169-3536}
\urldef\tempurl\url{https://doi.org/10.1109/ACCESS.2018.2887201}
\showDOI{\tempurl}


\bibitem[\protect\citeauthoryear{Smith}{Smith}{2018}]        {smith2018facerecognition}
\bibfield{author}{\bibinfo{person}{Brad Smith}.}
  \bibinfo{year}{2018}\natexlab{}.
\newblock \bibinfo{booktitle}{\emph{{Facial Recognition: It's Time for
  Action}}}.
\newblock
\urldef\tempurl\url{https://blogs.microsoft.com/on-the-issues/2018/12/06/facial-recognition-its-time-for-action/}
\showURL{\tempurl}


\bibitem[\protect\citeauthoryear{Tram{\`e}r, Zhang, Juels, Reiter, and
  Ristenpart}{Tram{\`e}r et~al\mbox{.}}{2016}]        {tramer2016stealing}
\bibfield{author}{\bibinfo{person}{Florian Tram{\`e}r}, \bibinfo{person}{Fan
  Zhang}, \bibinfo{person}{Ari Juels}, \bibinfo{person}{Michael~K Reiter},
  {and} \bibinfo{person}{Thomas Ristenpart}.} \bibinfo{year}{2016}\natexlab{}.
\newblock \showarticletitle{Stealing Machine Learning Models via Prediction
  APIs}. In \bibinfo{booktitle}{\emph{25th USENIX Security Symposium}}.
  \bibinfo{pages}{601--618}.
\newblock


\bibitem[\protect\citeauthoryear{Union}{Union}{2000}]        {ecomdirective}
\bibfield{author}{\bibinfo{person}{European Union}.}
  \bibinfo{year}{2000}\natexlab{}.
\newblock \bibinfo{title}{Directive 2000/31/EC of the European Parliament and
  of the Council of 8 June 2000 on certain legal aspects of information society
  services, in particular electronic commerce, in the Internal Market
  (E-Commerce Directive) OJ L 178}.
\newblock
\newblock


\bibitem[\protect\citeauthoryear{Union}{Union}{2016}]        {gdpr}
\bibfield{author}{\bibinfo{person}{European Union}.}
  \bibinfo{year}{2016}\natexlab{}.
\newblock \bibinfo{title}{Regulation (EU) 2016/679 of the European Parliament
  and of the Council of 27 April 2016 on the protection of natural persons with
  regard to the processing of personal data and on the free movement of such
  data, and repealing Directive 95/46/EC (General Data Protection Regulation)
  OJ L 119}.
\newblock
\newblock


\bibitem[\protect\citeauthoryear{Union}{Union}{2019}]        {copyright}
\bibfield{author}{\bibinfo{person}{European Union}.}
  \bibinfo{year}{2019}\natexlab{}.
\newblock \bibinfo{title}{Directive (EU) 2019/790 of the European Parliament
  and of the Council of 17 April 2019 on copyright and related rights in the
  Digital Single Market and amending Directives 96/9/EC and 2001/29/EC OJ L
  130}.
\newblock
\newblock


\bibitem[\protect\citeauthoryear{Veale, Binns, and Edwards}{Veale
  et~al\mbox{.}}{2018}]        {inversion}
\bibfield{author}{\bibinfo{person}{Michael Veale}, \bibinfo{person}{Reuben
  Binns}, {and} \bibinfo{person}{Lilian Edwards}.}
  \bibinfo{year}{2018}\natexlab{}.
\newblock \showarticletitle{Algorithms that remember: model inversion attacks
  and data protection law}.
\newblock \bibinfo{journal}{\emph{Philosophical Transactions of the Royal
  Society A: Mathematical, Physical and Engineering Sciences}}
  \bibinfo{volume}{376}, \bibinfo{number}{2133} (\bibinfo{year}{2018}),
  \bibinfo{pages}{20180083}.
\newblock
\urldef\tempurl\url{https://doi.org/10.1098/rsta.2018.0083}
\showDOI{\tempurl}


\bibitem[\protect\citeauthoryear{Walker}{Walker}{2018}]        {kent2018gooleai}
\bibfield{author}{\bibinfo{person}{Kent Walker}.}
  \bibinfo{year}{2018}\natexlab{}.
\newblock \bibinfo{booktitle}{\emph{{AI for Social Good in Asia Pacific}}}.
\newblock
\urldef\tempurl\url{https://www.blog.google/around-the-globe/google-asia/ai-social-good-asia-pacific/amp/}
\showURL{\tempurl}


\bibitem[\protect\citeauthoryear{Whittaker}{Whittaker}{2018}]        {whittaker_2018}
\bibfield{author}{\bibinfo{person}{Sarah Whittaker}.}
  \bibinfo{year}{2018}\natexlab{}.
\newblock \bibinfo{title}{What's The Big Deal? The Controversy on Google's AI
  and Pentagon Drones}.
\newblock
\newblock
\urldef\tempurl\url{https://dronebelow.com/2018/03/08/whats-the-big-deal-the-controversy-on-googles-ai-and-pentagon-drones/}
\showURL{\tempurl}


\bibitem[\protect\citeauthoryear{Zak}{Zak}{2019}]        {doffman2019hong-kong}
\bibfield{author}{\bibinfo{person}{Doffman Zak}.}
  \bibinfo{year}{2019}\natexlab{}.
\newblock \bibinfo{booktitle}{\emph{{Hong Kong Exposes Both Sides Of China's
  Relentless Facial Recognition Machine}}}.
\newblock
\urldef\tempurl\url{https://www.forbes.com/sites/zakdoffman/2019/08/26/hong-kong-exposes-both-sides-of-chinas-relentless-facial-recognition-machine/}
\showURL{\tempurl}


\bibitem[\protect\citeauthoryear{Zeng, Lu, Sun, and Tian}{Zeng
  et~al\mbox{.}}{2019}]        {zeng2019responsible}
\bibfield{author}{\bibinfo{person}{Yi Zeng}, \bibinfo{person}{Enmeng Lu},
  \bibinfo{person}{Yinqian Sun}, {and} \bibinfo{person}{Ruochen Tian}.}
  \bibinfo{year}{2019}\natexlab{}.
\newblock \showarticletitle{{Responsible Facial Recognition and Beyond}}.
\newblock \bibinfo{journal}{\emph{arXiv:1909.12935}} (\bibinfo{year}{2019}).
\newblock


\bibitem[\protect\citeauthoryear{Zhou, Pan, Wang, and Vasilakos}{Zhou
  et~al\mbox{.}}{2017}]        {Zhou_Pan_Wang_Vasilakos_2017}
\bibfield{author}{\bibinfo{person}{Lina Zhou}, \bibinfo{person}{Shimei Pan},
  \bibinfo{person}{Jianwu Wang}, {and} \bibinfo{person}{Athanasios~V.
  Vasilakos}.} \bibinfo{year}{2017}\natexlab{}.
\newblock \showarticletitle{Machine Learning on Big Data: Opportunities and
  Challenges}.
\newblock \bibinfo{journal}{\emph{Neurocomputing}} \bibinfo{volume}{237},
  \bibinfo{number}{December 2016} (\bibinfo{year}{2017}),
  \bibinfo{pages}{350--361}.
\newblock


\end{thebibliography}

\end{document}